
\font\sixrm=cmr6 at 6pt
\font\eightrm=cmr8 at 8pt

\font\seventeenrm=cmr17 at 17pt
\font\twentyonerm=cmr17 at 21pt

\font\ss=cmss10

\font\csc=cmcsc10

\font\twelvecal=cmsy10 at 12pt

\font\twelvemath=cmmi12

\font\seventeenbold=cmbx7 at 17pt

\font\fively=lasy5
\font\sevenly=lasy7
\font\tenly=lasy10

\textfont10=\tenly
\scriptfont10=\sevenly
\scriptscriptfont10=\fively
\magnification=1200
\parskip=10pt
\parindent=20pt
\def\today{\ifcase\month\or January\or February\or March\or April\or May\or
June
       \or July\or August\or September\or October\or November\or December\fi
       \space\number\day, \number\year}

\def\title#1{\footline={\ifnum\pageno<2\hfil
       \else\hss\tenrm\folio\hss\fi}\vskip1truein\centerline{{#1}
       \footnote{\raise1ex\hbox{*}}{\eightrm Supported in part
       by the Robert A. Welch Foundation and N.S.F. Grants
       PHY-880637 and\break PHY-8605978.}}}

\def\newpage{\vfill\eject}
\def\abstract#1{\centerline{\bf ABSTRACT}\vskip.2truein{\narrower\noindent#1
       \smallskip}}
\def\acknowledgements{\noindent\line{\bf Acknowledgements\hfill}\nobreak
    \vskip.1truein\nobreak\noindent\ignorespaces}
\def\runninghead#1#2{\voffset=2\baselineskip\nopagenumbers
       \headline={\ifodd\pageno\rightheadline\else \leftheadline\fi}
       \def\rightheadline{{\sl#1}\hfill{\rm\folio}}
       \def\leftheadline{{\rm\folio}\hfill{\sl#2}}}
\def\SS{\mathhexbox278}

\newcount\footnoteno
\def\Footnote#1{\advance\footnoteno by 1
                \let\SF=\empty
                \ifhmode\edef\SF{\spacefactor=\the\spacefactor}\/\fi
                $^{\the\footnoteno}$\ignorespaces
                \SF\vfootnote{$^{\the\footnoteno}$}{#1}}

\def\place#1#2#3{\vbox to0pt{\kern-\parskip\kern-7pt
                             \kern-#2truein\hbox{\kern#1truein #3}
                             \vss}\nointerlineskip}
\def\figurecaption#1#2{\kern.75truein\vbox{\hsize=5truein\noindent{\bf Figure
    \figlabel{#1}:} #2}}
\def\tablecaption#1#2{\kern.75truein\lower12truept\hbox{\vbox{\hsize=5truein
    \noindent{\bf Table\hskip5truept\tablabel{#1}:} #2}}}
\def\boxed#1{\lower3pt\hbox{
                       \vbox{\hrule\hbox{\vrule

\vbox{\kern2pt\hbox{\kern3pt#1\kern3pt}\kern3pt}\vrule}
                         \hrule}}}

\def\g{\gamma}
\def\d{\delta}\def\D{\Delta}

\def\l{\lambda}

\def\ca#1{\relax\ifmmode {{\cal #1}}\else $\cal #1$\fi}

\def\calb{{\cal B}}

\def\calm{{\cal M}}

\def\inbar{\vrule height1.5ex width.4pt depth0pt}
\def\IB{\relax{\rm I\kern-.18em B}}
\def\IC{\relax\hbox{\kern.25em$\inbar\kern-.3em{\rm C}$}}
\def\ID{\relax{\rm I\kern-.18em D}}
\def\IE{\relax{\rm I\kern-.18em E}}
\def\IF{\relax{\rm I\kern-.18em F}}
\def\IG{\relax\hbox{\kern.25em$\inbar\kern-.3em{\rm G}$}}
\def\IH{\relax{\rm I\kern-.18em H}}
\def\II{\relax{\rm I\kern-.18em I}}
\def\IK{\relax{\rm I\kern-.18em K}}
\def\IL{\relax{\rm I\kern-.18em L}}
\def\IM{\relax{\rm I\kern-.18em M}}
\def\IN{\relax{\rm I\kern-.18em N}}
\def\IO{\relax\hbox{\kern.25em$\inbar\kern-.3em{\rm O}$}}
\def\IP{\relax{\rm I\kern-.18em P}}
\def\IQ{\relax\hbox{\kern.25em$\inbar\kern-.3em{\rm Q}$}}
\def\IR{\relax{\rm I\kern-.18em R}}
\def\IZ{\relax\ifmmode\hbox{\ss Z\kern-.4em Z}\else{\ss Z\kern-.4em Z}\fi}
\def\IGa{\relax{\rm I}\kern-.18em\Gamma}
\def\IPi{\relax{\rm I}\kern-.18em\Pi}
\def\ITh{\relax\hbox{\kern.25em$\inbar\kern-.3em\Theta$}}
\def\IOm{\relax\thinspace\inbar\kern1.95pt\inbar\kern-5.525pt\Omega}


\def\define{\buildrel\rm def\over =}

\def\cy{Calabi--Yau}
\def\cym{Calabi--Yau manifold}
\def\cys{Calabi--Yau manifolds}

\def\H#1#2{\relax\ifmmode {H^{#1#2}}\else $H^{#1 #2}$\fi}
\def\M{\relax\ifmmode{\calm}\else $\calm$\fi}

\def\Bigcheck{\lower3.8pt\hbox{\smash{\hbox{{\twentyonerm \v{}}}}}}
\def\bigboldcheck{\smash{\hbox{{\seventeenbold\v{}}}}}

\def\Bighat{\lower3.8pt\hbox{\smash{\hbox{{\twentyonerm \^{}}}}}}

\def\Msharp{\relax\ifmmode{\calm^\sharp}\else $\smash{\calm^\sharp}$\fi}
\def\Mflat{\relax\ifmmode{\calm^\flat}\else $\smash{\calm^\flat}$\fi}
\def\preMcheck{\kern2pt\hbox{\Bigcheck\kern-12pt{$\cal M$}}}
\def\Mcheck{\relax\ifmmode\preMcheck\else $\preMcheck$\fi}
\def\preMhat{\kern2pt\hbox{\Bighat\kern-12pt{$\cal M$}}}
\def\Mhat{\relax\ifmmode\preMhat\else $\preMhat$\fi}

\def\Bsharp{\relax\ifmmode{\calb^\sharp}\else $\calb^\sharp$\fi}
\def\Bflat{\relax\ifmmode{\calb^\flat}\else $\calb^\flat$ \fi}
\def\preBcheck{\hbox{\Bigcheck\kern-9pt{$\cal B$}}}
\def\Bcheck{\relax\ifmmode\preBcheck\else $\preBcheck$\fi}
\def\preBhat{\hbox{\Bighat\kern-9pt{$\cal B$}}}
\def\Bhat{\relax\ifmmode\preBhat\else $\preBhat$\fi}

\def\figBcheck{\kern3pt\hbox{\raise1pt\hbox{\bigboldcheck}\kern-11pt
    {\twelvecal B}}}
\def\figBsharp{{\twelvecal B}\raise5pt\hbox{$\twelvemath\sharp$}}
\def\figBflat{{\twelvecal B}\raise5pt\hbox{$\twelvemath\flat$}}

\def\gcheck{\hbox{\lower2.5pt\hbox{\Bigcheck}\kern-8pt$\g$}}
\def\lhat{\hbox{\raise.5pt\hbox{\Bighat}\kern-8pt$\l$}}

\def\Fcheck{\kern2pt\hbox{\raise1pt\hbox{\Bigcheck}\kern-10pt{$\cal F$}}}
\def\Fhat{\kern2pt\hbox{\raise1pt\hbox{\Bighat}\kern-10pt{$\cal F$}}}

\def\cp#1{\relax\ifmmode {\IP\kern-2pt{}_{#1}}\else $\IP\kern-2pt{}_{#1}$\fi}
\def\h#1#2{\relax\ifmmode {b_{#1#2}}\else $b_{#1#2}$\fi}

\def\frac#1#2{{#1\over #2}}

\def\cone{\relax\thinspace\hbox{$<\kern-.8em{)}$}}
\mathchardef\mho"0A30

\def\-{\hphantom{-}}


\def\npb#1{Nucl.\ Phys.\ {\bf B#1}}


\newif\ifproofmode
\proofmodefalse

\newif\ifforwardreference
\forwardreferencefalse

\newif\ifchapternumbers
\chapternumbersfalse

\newif\ifcontinuousnumbering
\continuousnumberingfalse

\newif\iffigurechapternumbers
\figurechapternumbersfalse

\newif\ifcontinuousfigurenumbering
\continuousfigurenumberingfalse

\newif\iftablechapternumbers
\tablechapternumbersfalse

\newif\ifcontinuoustablenumbering
\continuoustablenumberingfalse

\font\eqsixrm=cmr6

\font\sixrm=cmr6 at 6pt

\def\marginstyle{\eqsixrm}

\newtoks\chapletter
\newcount\chapno
\newcount\eqlabelno
\newcount\figureno
\newcount\tableno

\chapno=0
\eqlabelno=0
\figureno=0
\tableno=0

\def\chapfolio{\ifnum\chapno>0 \the\chapno\else\the\chapletter\fi}

\def\bumpchapno{\ifnum\chapno>-1 \global\advance\chapno by 1
\else\global\advance\chapno by -1 \setletter\chapno\fi
\ifcontinuousnumbering\else\global\eqlabelno=0 \fi
\ifcontinuousfigurenumbering\else\global\figureno=0 \fi
\ifcontinuoustablenumbering\else\global\tableno=0 \fi}

\def\setletter#1{\ifcase-#1{}\or{}%
\or\global\chapletter={A}%
\or\global\chapletter={B}%
\or\global\chapletter={C}%
\or\global\chapletter={D}%
\or\global\chapletter={E}%
\or\global\chapletter={F}%
\or\global\chapletter={G}%
\or\global\chapletter={H}%
\or\global\chapletter={I}%
\or\global\chapletter={J}%
\or\global\chapletter={K}%
\or\global\chapletter={L}%
\or\global\chapletter={M}%
\or\global\chapletter={N}%
\or\global\chapletter={O}%
\or\global\chapletter={P}%
\or\global\chapletter={Q}%
\or\global\chapletter={R}%
\or\global\chapletter={S}%
\or\global\chapletter={T}%
\or\global\chapletter={U}%
\or\global\chapletter={V}%
\or\global\chapletter={W}%
\or\global\chapletter={X}%
\or\global\chapletter={Y}%
\or\global\chapletter={Z}\fi}

\def\tempsetletter#1{\ifcase-#1{}\or{}%
\or\global\chapletter={A}%
\or\global\chapletter={B}%
\or\global\chapletter={C}%
\or\global\chapletter={D}%
\or\global\chapletter={E}%
\or\global\chapletter={F}%
\or\global\chapletter={G}%
\or\global\chapletter={H}%
\or\global\chapletter={I}%
\or\global\chapletter={J}%
\or\global\chapletter={K}%
\or\global\chapletter={L}%
\or\global\chapletter={M}%
\or\global\chapletter={N}%
\or\global\chapletter={O}%
\or\global\chapletter={P}%
\or\global\chapletter={Q}%
\or\global\chapletter={R}%
\or\global\chapletter={S}%
\or\global\chapletter={T}%
\or\global\chapletter={U}%
\or\global\chapletter={V}%
\or\global\chapletter={W}%
\or\global\chapletter={X}%
\or\global\chapletter={Y}%
\or\global\chapletter={Z}\fi}

\def\chapshow#1{\ifnum#1>0 \relax#1%
\else{\tempsetletter{\number#1}\chapno=#1\chapfolio}\fi}

\def\chaplabel#1{\bumpchapno\ifproofmode\ifforwardreference
\immediate\write1{\noexpand\expandafter\noexpand\def
\noexpand\csname CHAPLABEL#1\endcsname{\the\chapno}}\fi\fi
\global\expandafter\edef\csname CHAPLABEL#1\endcsname
{\the\chapno}\ifproofmode\llap{\hbox{\marginstyle #1\ }}\fi\chapfolio}

\def\eqnum{\global\advance\eqlabelno by 1
\eqno(\ifchapternumbers\chapfolio.\fi\the\eqlabelno)}

\def\eqlabel#1{\global\advance\eqlabelno by 1 \ifproofmode\ifforwardreference
\immediate\write1{\noexpand\expandafter\noexpand\def
\noexpand\csname EQLABEL#1\endcsname{\the\chapno.\the\eqlabelno?}}\fi\fi
\global\expandafter\edef\csname EQLABEL#1\endcsname
{\the\chapno.\the\eqlabelno?}\eqno(\ifchapternumbers\chapfolio.\fi
\the\eqlabelno)\ifproofmode\rlap{\hbox{\marginstyle #1}}\fi}

\def\eqalignnum{\global\advance\eqlabelno by 1
&(\ifchapternumbers\chapfolio.\fi\the\eqlabelno)}

\def\eqalignlabel#1{\global\advance\eqlabelno by 1 \ifproofmode
\ifforwardreference\immediate\write1{\noexpand\expandafter\noexpand\def
\noexpand\csname EQLABEL#1\endcsname{\the\chapno.\the\eqlabelno?}}\fi\fi
\global\expandafter\edef\csname EQLABEL#1\endcsname
{\the\chapno.\the\eqlabelno?}&(\ifchapternumbers\chapfolio.\fi
\the\eqlabelno)\ifproofmode\rlap{\hbox{\marginstyle #1}}\fi}

\def\eqref#1{\hbox{(\ifundefined{EQLABEL#1}***)\ifproofmode\ifforwardreference%
\else\write16{ ***Undefined Equation Reference #1*** }\fi
\else\write16{ ***Undefined Equation Reference #1*** }\fi
\else\edef\LABxx{\getlabel{EQLABEL#1}}%
\def\LAByy{\expandafter\stripchap\LABxx}\ifchapternumbers%
\chapshow{\LAByy}.\expandafter\stripeq\LABxx%
\else\ifnum\number\LAByy=\chapno\relax\expandafter\stripeq\LABxx%
\else\chapshow{\LAByy}.\expandafter\stripeq\LABxx\fi\fi)\fi}%
\ifproofmode\write2{Equation #1}\fi}

\def\fignum{\global\advance\figureno by 1
\relax\iffigurechapternumbers\chapfolio.\fi\the\figureno}

\def\figlabel#1{\global\advance\figureno by 1
\relax\ifproofmode\ifforwardreference
\immediate\write1{\noexpand\expandafter\noexpand\def
\noexpand\csname FIGLABEL#1\endcsname{\the\chapno.\the\figureno?}}\fi\fi
\global\expandafter\edef\csname FIGLABEL#1\endcsname
{\the\chapno.\the\figureno?}\iffigurechapternumbers\chapfolio.\fi
\ifproofmode\llap{\hbox{\marginstyle#1
\kern1.2truein}}\relax\fi\the\figureno}

\def\figref#1{\hbox{\ifundefined{FIGLABEL#1}!!!!\ifproofmode%
\ifforwardreference%
\else\write16{ ***Undefined Figure Reference #1*** }\fi
\else\write16{ ***Undefined Figure Reference #1*** }\fi
\else\edef\LABxx{\getlabel{FIGLABEL#1}}%
\def\LAByy{\expandafter\stripchap\LABxx}\iffigurechapternumbers%
\chapshow{\LAByy}.\expandafter\stripeq\LABxx%
\else\ifnum \number\LAByy=\chapno\relax\expandafter\stripeq\LABxx%
\else\chapshow{\LAByy}.\expandafter\stripeq\LABxx\fi\fi\fi}%
\ifproofmode\write2{Figure #1}\fi}

\def\tabnum{\global\advance\tableno by 1
\relax\iftablechapternumbers\chapfolio.\fi\the\tableno}

\def\tablabel#1{\global\advance\tableno by 1
\relax\ifproofmode\ifforwardreference
\immediate\write1{\noexpand\expandafter\noexpand\def
\noexpand\csname TABLABEL#1\endcsname{\the\chapno.\the\tableno?}}\fi\fi
\global\expandafter\edef\csname TABLABEL#1\endcsname
{\the\chapno.\the\tableno?}\iftablechapternumbers\chapfolio.\fi
\ifproofmode\llap{\hbox{\marginstyle#1
\kern1.2truein}}\relax\fi\the\tableno}

\def\tabref#1{\hbox{\ifundefined{TABLABEL#1}!!!!\ifproofmode%
\ifforwardreference%
\else\write16{ ***Undefined Table Reference #1*** }\fi
\else\write16{ ***Undefined Table Reference #1*** }\fi
\else\edef\LABtt{\getlabel{TABLABEL#1}}%
\def\LABTT{\expandafter\stripchap\LABtt}\iftablechapternumbers%
\chapshow{\LABTT}.\expandafter\stripeq\LABtt%
\else\ifnum\number\LABTT=\chapno\relax\expandafter\stripeq\LABtt%
\else\chapshow{\LABTT}.\expandafter\stripeq\LABtt\fi\fi\fi}%
\ifproofmode\write2{Table#1}\fi}

\newdimen\sectionskip     \sectionskip=20truept
\newcount\sectno
\def\section#1#2{\sectno=0 \null\vskip\sectionskip
    \centerline{\chaplabel{#1}.~~{\bf#2}}\nobreak\vskip.2truein
    \noindent\ignorespaces}

\def\advancesectno{\global\advance\sectno by 1}
\def\sectfolio{\number\sectno}
\def\subsection#1{\goodbreak\advancesectno\null\vskip10pt
                  \noindent\chapfolio.~\sectfolio.~{\bf #1}
                  \nobreak\vskip.05truein\noindent\ignorespaces}

\def\uttg#1{\null\vskip.1truein
    \ifproofmode \line{\hfill{\bf Draft}:
    UTTG--{#1}--\number\year}\line{\hfill\today}
    \else \line{\hfill UTTG--{#1}--\number\year}
    \line{\hfill\ifcase\month\or January\or February\or March\or April\or
May\or June
    \or July\or August\or September\or October\or November\or December\fi
    \space\number\year}\fi}

\def\contents{\noindent
   {\bf Contents\Z}\nobreak\vskip.05truein\noindent\ignorespaces}

\def\getlabel#1{\csname#1\endcsname}
\def\ifundefined#1{\expandafter\ifx\csname#1\endcsname\relax}
\def\stripchap#1.#2?{#1}
\def\stripeq#1.#2?{#2}

%
\catcode`@=11 
\def\space@ver#1{\let\@sf=\empty\ifmmode#1\else\ifhmode%
\edef\@sf{\spacefactor=\the\spacefactor}\unskip${}#1$\relax\fi\fi}
\newcount\referencecount     \referencecount=0
\newif\ifreferenceopen       \newwrite\referencewrite
\newtoks\rw@toks
\def\refmark#1{\relax[#1]}
\def\refend{\refmark{\number\referencecount}}
\newcount\lastrefsbegincount \lastrefsbegincount=0
\def\refsend{\refmark{\count255=\referencecount%
\advance\count255 by -\lastrefsbegincount%
\ifcase\count255 \number\referencecount%
\or\number\lastrefsbegincount,\number\referencecount%
\else\number\lastrefsbegincount-\number\referencecount\fi}}
\def\refch@ck{\chardef\rw@write=\referencewrite
\ifreferenceopen\else\referenceopentrue
\immediate\openout\referencewrite=referenc.texauxil \fi}
%
{\catcode`\^^M=\active 
  \gdef\obeyendofline{\catcode`\^^M\active \let^^M\ }}%
%
{\catcode`\^^M=\active 
  \gdef\ignoreendofline{\catcode`\^^M=5}}
{\obeyendofline\gdef\rw@start#1{\def\t@st{#1}\ifx\t@st\blankend%
\endgroup\@sf\relax\else\ifx\t@st\bl@nkend\endgroup\@sf\relax%
\else\rw@begin#1
\backtotext
\fi\fi}}
{\obeyendofline\gdef\rw@begin#1
{\def\n@xt{#1}\rw@toks={#1}\relax%
\rw@next}}
\def\blankend{}
{\obeylines\gdef\bl@nkend{
}}
\newif\iffirstrefline  \firstreflinetrue
\def\rwr@teswitch{\ifx\n@xt\blankend\let\n@xt=\rw@begin%
\else\iffirstrefline\global\firstreflinefalse%
\immediate\write\rw@write{\noexpand\obeyendofline\the\rw@toks}%
\let\n@xt=\rw@begin%
\else\ifx\n@xt\rw@@d \def\n@xt{\immediate\write\rw@write{%
\noexpand\ignoreendofline}\endgroup\@sf}%
\else\immediate\write\rw@write{\the\rw@toks}%
\let\n@xt=\rw@begin\fi\fi\fi}
\def\rw@next{\rwr@teswitch\n@xt}
\def\rw@@d{\backtotext} \let\rw@end=\relax
\let\backtotext=\relax

\newdimen\refindent     \refindent=30pt
\def\Textindent#1{\noindent\llap{#1\enspace}\ignorespaces}
\def\refitem#1{\par\hangafter=0 \hangindent=\refindent\Textindent{#1}}
\def\REFNUM#1{\space@ver{}\refch@ck\firstreflinetrue%
\global\advance\referencecount by 1 \xdef#1{\the\referencecount}}
\def\refnum#1{\space@ver{}\refch@ck\firstreflinetrue%
\global\advance\referencecount by 1\xdef#1{\the\referencecount}\refend}

\def\REF#1{\REFNUM#1%
\immediate\write\referencewrite{%
\noexpand\refitem{#1.}}%
\begingroup\obeyendofline\rw@start}
\def\ref{\refnum\?%
\immediate\write\referencewrite{\noexpand\refitem{\?.}}%
\begingroup\obeyendofline\rw@start}
\def\Ref#1{\refnum#1%
\immediate\write\referencewrite{\noexpand\refitem{#1.}}%
\begingroup\obeyendofline\rw@start}
\def\REFS#1{\REFNUM#1\global\lastrefsbegincount=\referencecount%
\immediate\write\referencewrite{\noexpand\refitem{#1.}}%
\begingroup\obeyendofline\rw@start}

\def\cite#1{\refmark#1}
\catcode`@=12 
%
\proofmodefalse
\chapternumberstrue
\ifproofmode
\immediate\openout2=allcrossreferfile \fi
\ifforwardreference\input labelfile
\ifproofmode\immediate\openout1=labelfile \fi\fi

\def\vone{{\bf 1}}
\def\subsection#1{\goodbreak\advancesectno\null\vskip10pt
                  \noindent{\it \chapfolio.\sectfolio.~#1}
                  \nobreak\vskip.05truein\noindent\ignorespaces}
\def\contents{\line{{\bf Contents}\hfill}\nobreak\vskip.05truein\noindent
              \ignorespaces}

\def\cropen#1{\crcr\noalign{\vskip #1}}

%
\nopagenumbers\pageno=-1
\null\vskip-20pt
\rightline{\eightrm UTTG-20-95}\vskip-3pt
\rightline{\eightrm hepth/9511230}\vskip-3pt
\rightline{\eightrm 29 November 1995}
\vskip1truein
\centerline{\seventeenrm On the Connectedness of the Moduli Space}
\vskip5pt
\centerline{\seventeenrm of Calabi--Yau Manifolds}
\vskip.7truein
\centerline{\csc
A.C.~Avram$\,^1$ ,~
P.~Candelas$\,^2$ ,~
D.~Jan\v{c}i\'{c}$\,^3$  ~ and~
M.~Mandelberg$\,^4$}
\footnote{}{\hbox to 3in{\eightrm $^1\,$Email: alex@physics.utexas.edu\hfill}
\eightrm $^2\,$Email: candelas@physics.utexas.edu}
\footnote{}{\hbox to 3in{\eightrm $^3\,$Email: jancic@physics.utexas.edu\hfill}
\eightrm $^4\,$Email: isaac@physics.utexas.edu}
\vskip.7truein
\centerline{\it Theory Group}
\centerline{\it Department of Physics}
\centerline{\it University of Texas}
\centerline{\it Austin, TX 78712}
\vskip1.4truein
\vbox{\centerline{\bf ABSTRACT}
\vskip.2truein
\vbox{\baselineskip=13pt\noindent We show that the moduli space of all \cys\
that can be realized as hypersurfaces described by a transverse polynomial in a
four dimensional
weighted projective space, is connected. This is achieved by exploiting
techniques of toric geometry and the construction of Batyrev that relate \cys\
to reflexive polyhedra. Taken
together with the previously known fact that the moduli space of all CICY's is
connected,
and is moreover connected to the moduli space of the present class of \cys\
(since the
quintic threefold $\IP_4 \lbrack 5 \rbrack$ is both CICY and a hypersurface in
a weighted
$\IP_4$), this strongly suggests that the moduli space of all simply connected
\cys\ is
connected. It is of interest that singular \cys\ corresponding to the points in
which the moduli spaces meet are often, for the present class, more singular
than the conifolds that connect the moduli spaces of CICY's. }}
\vfill\eject
\pageno=1\footline={\rm\hfil\folio\hfil}
\baselineskip=18pt plus 1pt minus 1pt

%
\REF{\rReid}{M.~Reid, Math.\ Ann.\ {\bf 278} (1987) 329.}
\REF{\rCDLS}{P.~Candelas, A.M.~Dale, C.A.~L\"utken, R.~Schimmrigk,\hfill\break
       \npb{298}~(1988)~493.}
\REF{\rRolling}{P.~Candelas, P.S.~Green and T.~H\"ubsch,
      \npb{330} (1990) 49.}
\REF{\rAGM}{P.~S.Aspinwall, B.R.~Greene and D.R.~Morrison,\hfil\break
Int.~Math.~Res.~Notices (1993)~ 319, alg-geom/9309007.}
\REF{\rBeast}{T.~H\"ubsch, {\it \cy\ Manifolds--A Bestiary for
       Physicists},\hfil\break (World Scientific, Singapore, 1992).}
\REF{\rTris}{T.~H\"ubsch, Commun.~Math.~Phys.~{\bf 108}~(1987)~291.}
\REF{\rPGTH}{P.~Green, T.~H\"ubsch, Commun.~Math.~Phys.~{\bf 109}~(1987)~99.}
\REF{\rHeC}{A.~He and P.~Candelas, ~Commun.~Math. Phys.~{\bf 135}~193~(1990).}
\REF{\rCLS}{P.~Candelas, M.~Lynker and R.~Schimmrigk, \npb{341}~(1990)~383.}
\REF{\rKS}{A.~Klemm, R.~Schimmrigk, \npb{411}~(1994)~559.}
\REF{\rMaxSkI}{M.~Kreuzer, H.~Skarke, \npb{388}~(1992)~113.}
\REF{\rLS}{M.~Lynker and R.~Schimmrigk, ``Conifold Transitions and
Mirror Symmetries''\hfill\break hep-th/9511058}
\REF{\rBB}{V.~Batyrev and B.~Borisov, ``Dual Cones and Mirror Symmetry for
Generalised Calabi-Yau Manifolds''~alg-geom/9402002.}
\REF{\rB}{V.~Batyrev, Duke Math. Journal, Vol 69, No 2, (1993)
      349, ~alg-geom/9310003.}
\REF{\rCDK}{P.~Candelas, Xenia~de la Ossa and
S.~Katz, \npb{450}~(1995)~267,\hfil\break hep-th/9412117.}
\REF{\rBKK}{P.~Berglund, S.~Katz and A.~Klemm, ``Mirror Symmetry and
the Moduli Space for Generic Hypersurfaces in Toric Varieties'',
hep-th/9506091.}
\REF{\rHaya}{Y.~Hayakawa, ``Degeneration of Calabi--Yau Manifold
With Weil--Peterson  Metric'', ~alg-geom/9507016.}
\REF{\rStro}{A.~Strominger, ``Massless Black Holes and Conifold in String
Theory'', \hfill\break ~hep-th/9504090.}
\REF{\rBMS}{B.~Greene, D.~R.~Morrison and A.~Strominger, \npb{451}~(1995)~109,
\hfill\break hep-th/9504145.}
\REF{\rCGGK}{T.~Chiang, B.~Greene, M.~Gross and Y.~Kanter, ``Black
Hole Condensation and the Web of Calabi--Yau Manifolds'', hep-th/9511204.}
{\baselineskip=13pt
\contents
\vskip15pt
\item{1.~}Introduction
\bigskip
\item{2.~}Nesting of Reflexive Polyhedra
\bigskip
\item{3.~}Calabi-Yau Manifolds in Projective Varieties
\itemitem{\it 3.1~}{\it Generalities}
\itemitem{\it 3.2~}{\it Connectivity of Moduli Spaces}
\itemitem{\it 3.3~}{\it Illustration of the Method}
\bigskip
\item{4.~}The Computation
\bigskip
\item{A.~}Appendix: The $n$-Vertex Irreducible RP's that Connect the
Manifolds}
\newpage
\section{intro}{Introduction}
It has been known for some time \cite{{\rReid,\rCDLS,\rRolling,\rAGM}} that the
moduli spaces of
some \cys\ meet along boundary components that correspond to certain singular
manifolds.  Differently put: certain singular \cys\ can be approached as limits
of deformation classes of topologically distinct manifolds.  In
\cite{\rReid} M. Reid
made a bold conjecture that the parameter space of threefolds with vanishing
first Chern class is connected.  In ~\cite{{\rRolling,\rBeast}} it was shown
that the parameter space
of all CICY's is connected.  CICY's are complete intersection Calabi-Yau
manifolds~\cite{{\rCDLS,\rBeast,\rTris,\rPGTH}}: \cys\ that can be realized as
complete intersections of polynomials
defined on products of projective spaces.  The class of CICY's comprises
several
thousand~\cite{\rHeC} topologically distinct manifolds corresponding to
some 250 pairs of values for the Hodge numbers ($h_{11},h_{21}$) and with Euler
numbers in the range $-200\leq\chi\leq 0$.  At the time of \cite{\rRolling}
this was the largest class of \cys\ that could be systematically constructed.
Since then
another, perhaps larger, class has been studied. These are manifolds that
can be realized by a transverse polynomial (a polynomial $p$ such that $p$ and
$dp$ do not
simultaneously vanish) in \Footnote{Strictly speaking, the
\cys\ are embedded in blow ups of $\IP_4^{\bf{k}}$} $\IP_4^{\bf{k}}$ , a
weighted $\IP_4$, with weights
${\bf k}=(k_1,k_2,k_3,k_4,k_5)$~\cite{{\rCLS,\rKS,\rMaxSkI}}.
The authors of
\cite{{\rKS,\rMaxSkI}} constructed a list ${\cal L}$ of 7555 weight vectors
$\bf{k}$ corresponding to these manifolds.  These weight vectors do not all
lead
to distinct CY manifolds though there are roughly 2,500 distinct pairs of Hodge
numbers. The Euler numbers of this class of manifolds lie in the range
$-960\leq\chi\leq 960$.  To our knowledge no
\cys\ are known with Euler numbers outside this range. In virtue of this, we
considered it of interest to ask whether the moduli
spaces of all the manifolds of the list $\cal L$
are connected together. The result that we report in this paper is that they
are.

Reid conjectured that the moduli spaces of all \cys\ are connected via
conifolds \cite{\rRolling}, though it would be in accord with the conjecture
for the
connection to be through non-K\"ahler manifolds with vanishing first
Chern class. We have found that, at least within our limited class,
it is not neccessary to leave the family of K\"ahler \cys\ in order to
show that the moduli spaces form a connected web. Though the singular manifolds
that connect the moduli spaces are in many cases more singular than conifolds.
Since the list contains the quintic threefold, $\IP_4[5]$, which is also a
CICY, it follows that these moduli spaces are connected also to the web of
CICY's.

For the case of CICY's it was possible to show that the moduli spaces
of the manifolds were connected by means of an analytic argument.
Each CICY is specified by a degree matrix and the authors of \cite{\rPGTH} have
shown that a CICY exists for each such matrix.
A conifold transition corresponds to a certain operation on the degree
matrix and it is straightforward to see that one can transform any two
given degree matrices into one another by a sequence of such moves.
The analogous argument for weighted CICY's fails however, at least in
its naive form, since there are degree matrices for which the
corresponding weighted CICY has terminal singularities (which can not
be resolved to give a manifold with $c_1 = 0$).
Although there is no general proof that the moduli spaces are connected
for weighted CICY's it is clear \cite{\rLS} that many weighted CICY's
are connected to the web of CICY's and weighted~$\IP_4$'s.

Our investigation relies heavily on the toric construction of \cys\ due to
Batyrev~\cite{{\rBB,\rB}}. For a manifold \ca{M}\ of a degree $d$, (defined as
the vanishing locus of a
polynomial $p$ of a degree $d$ in a weighted $\IP_4$) denote by $(x_1, \ldots,
x_5)$ the
homogeneous coordinates of the projective space and by ${\bf x}^{\bf m}$ the
monomial
$x_1^{m_1}\ldots x_5^{m_5}$. The set of all possible exponents ${\bf m}$
corresponding to a degree $d = k_1 + \ldots + k_5$ polynomial forms the Newton
polyhedron of $\ca{M}$. Batyrev observed that in many cases the Newton
polyhedron
of a manifold of the list $\ca{L}$ has a certain property termed reflexivity.
In ~\cite{\rCDK}
it was checked that in fact  all the manifolds of the list have this property.
Now a
converse obtains: if a polyhedron $\D$ is reflexive then a \cym\ $\ca{M}_\D$
may be constructed from $\D$. It may happen that a given reflexive polyhedron
contains a subsets of points that themselves form a reflexive polyhedron, $\d$.
When this happens the moduli spaces of $\ca{M}_\D$ and
$\ca{M}_\d$ intersect \cite{\rBKK}.
This is so because the polynomial of $\ca{M}_\d$
contains a subset of the monomials of that for $\ca{M}_\D$ and so can be
deformed continuously by letting some of the coefficients in the polynomial for
$\ca{M}_\D$ tend to zero. It follows also from a recent theorem of
Hayakawa~\cite{\rHaya} that the distance from the generic smooth manifold
$\ca{M}_\D$ or
$\ca{M}_\d$, to the singular manifolds $\ca{M}^\sharp$ (that correspond to the
intersection of the two moduli spaces) is~finite.

Recently, a possible explanation of a manner in which string theory unifies the
moduli spaces of many (possibly all) Calabi--Yau vacua was
discovered~\cite{{\rStro,\rBMS}}. It is argued that black hole condensation can
occur at conifold singularities in
the moduli space of type IIB  Calabi--Yau string vacua, and in some
cases this condensate
signals a smooth transition to a new, topologically different Calabi--Yau
vacuum. The extent to which it is possible to give a similar physical
interpretation to the more complicated
singularities that arise here is an interesting open question.

The paper is organized as follows. In \SS2 we try to motivate
intuitively  how we can deduce the connectivity of moduli
spaces from the nesting of polyhedra. In \SS3 we provide some of
the mathematical details to support this idea. After recalling some
facts about how reflexive polyhedra are constructed, we establish that
if one polyehedron contains another, then the moduli spaces of the
corresponding families of \cys\ are connected. We give also a concrete
example of this procedure. In \SS4 we provide some of the
details about the algorithm used to show the connectedness of the
moduli spaces associated with $\cal L$. Finally, in the appendix we
provide a list of polyhedra that realize the
connectedness as described in this article.

We are grateful to D.~Morrison for initially suggesting this problem to us.

Just prior to the submission of this article, we were made aware of similar
results reported in ~\cite{\rCGGK}.

\newpage

\section{p}{Nesting of Reflexive Polyhedra}
Newton polyhedra are important since they will allow us to describe how the
moduli spaces meet. Suppose that a Newton polyhedron $\D_1$ from $\cal L$
contains a reflexive polyhedron $\D_2$ as a subpolyhedron.
Then the rough idea
is that $\D_2$ is obtained from $\D_1$ by setting to zero certain coefficients
in the polynomial $p_1$ of $\D_1$. This is clearly a continuous operation. In
this way we see that the moduli space of the \cys\ corresponding to $\D_1$
intersects the moduli space of \cys\ corresponding to $\D_2$. If there is a
third polyhedron $\D_3$ that also contains $\D_2$
$$ \Delta_1 \supset \Delta_2 \subset \Delta_3
\eqlabel{inclusion}.$$
then the moduli space of the manifolds corresponding to $\Delta_3$ also
intersects the moduli space
of those corresponding to $\D_2$ and the three moduli spaces are connected.

The moduli spaces corresponding to all $\D$'s that contain a common
subpolyhedron are connected.
If we denote by $F_{\delta}$ the ``family'' of moduli spaces corresponding to
polyhedra that contain a given
subpolyheron $\delta$, and if any polyhedron of a
family $F_{\delta_1}$ has a subpolyhedron in common with any
polyhedron of a family $F_{\delta_2}$, then the two families
$F_{\delta_1}$ and $F_{\delta_2}$ are connected.

There are however some aspects of this process that need to be explained. The
apparent difficulty is that $\D_1$ corresponds to a family of hypersurfaces in
a weighted projective space, with weight vector ${\bf k}_1$ say, and $\D_2$ to
a hypersurface in a
 weighted projective space with a different weight ${\bf k}_2$, say. This
difficulty
is only apparent; the essential point is that, given a reflexive polyhedron
$\D$, a deformation class of \cys\ may be constructed from $\D$, such that the
generic manifold in the class is smooth. It is important that the reflexivity
of $\D$ is sufficient
for this to be true. $\D$ does not have to correspond to a member of $\ca{L}$.
This is
fortunate, since some of the polyhedra that we use to prove the connectivity of
the moduli space, do not belong to $\ca{L}$. The resolution of the apparent
difficulty concerning the different weight vectors, is in essence just a point
made above about the
 smoothness of the generic \cym\ corresponding to a reflexive polyhedron. By
setting to
 zero the coefficients that take us from $\D_1$ to $\D_2$, we obtain a singular
manifold
$\ca{M}^\sharp_1$, in $\IP^{{\bf k}_1}_4$. This same singular space can be
realized as the limit
of a hypersurface in $\IP^{{\bf k}_2}_4$. It is important to note that a family
of smooth \cys\
are in one--to--one correspondence with a dual\Footnote{The dual of a convex
polyhedron is defined in \SS3.1} pair of reflexive polyhedra, $(\D_1,
\nabla_1)$, say. There is, similarly, a family for $(\D_2, \nabla_2)$. Now,
$\D_1 \supset \D_2$ and it is important that  duality reverses the inclusion
for the dual polyhedra so that $\nabla_1 \subset \nabla_2$. The two families
meet in manifolds $\ca{M}^\sharp$, which may be considered to correspond to the
non--dual pair $(\D_2, \nabla_1)$.

We express this as a diagram:
$$
\matrix{\ca{M}_1:&~~~~~~\, (~\D_1,\nabla_1~)\cr
        ~~~~~~~~~&\downarrow\cr
        \ca{M}^\sharp:&~~~~~~\,(~\D_2,\nabla_1~)\cr
         ~~~~~~~~~&~~~~~~~~~~\, \uparrow\cr
         \ca{M}_2:&~~~~~~\,(~\D_2,\nabla_2~)\cr}~\eqlabel{*}$$
where the arrows denote specializations of either the polynomial corresponding
to $\ca{M}_1$ or the mirror polynomial corresponding to $\ca{M}_2$.

This diagram allows a nice interpretation. We may think of the singularization
$(\D_1, \nabla_1) \longrightarrow (\D_2, \nabla_1)$ as being due to the
specialization of the polynomial which forces the hypersurface to be singular.
Alternatively,
we can singularize via $(\D_2, \nabla_2) \longrightarrow (\D_2, \nabla_1)$.
We will see below that the process $\nabla_2 \longrightarrow \nabla_1$ can be
thought of
as a singularization of the embedding space. Speaking loosely, we can say that
we achieve the same effect by singularizing the hypersurface while leaving
alone the embedding space of one manifold, or by leaving the hypersurface alone
and
singularizing the embedding space of the other manifold.

The next section, which is rather technical, shows that the operations
that correspond to reversing the arrows in \eqref{*} render the
manifolds $\ca{M}_1$ and $\ca{M}_2$ smooth.
\newpage
\section{th}{Calabi-Yau Manifolds in Projective Varieties}
\vskip-40pt
\subsection{Generalities}
We consider \cys\ as hypersurfaces in toric varieties that are toric
deformations  of weighted projective spaces. These varieties, as well
as the relations between them, will be described in terms of toric geometry.
To this end we briefly review the toric construction of projective
varieties~\cite{{\rAGM,\rBB,\rB,\rCDK}}.

We start with a weighted projective space $\IP_4^{\bf k}$,  and consider the
family of homogeneous polynomials $p = p(x_1, \ldots , x_5)$ of degree
$d = \sum_{i=1}^{5} k_{i}$. As in \SS1, we associate a vector of exponents to
each
monomial and write $\bf x^{\bf m}$ for
$x_1^{m_1}x_2^{m_2}x_3^{m_3}x_4^{m_4}x_5^{m_5}$.
Thus, a general homogeneous polynomial of degree $d$ has the form:
$$p = \sum_{{{\bf k} \cdot {\bf m} = d,\atop m_i\ge0}}c_{\bf m}{\bf x}^{\bf
m}~.$$
Each degree vector $\bf m $ can be regarded as a point in $\IZ^5 \otimes{\IR}$,
and the convex hull of these points forms the Newton polyhedron $
\Delta({\bf k})$ of $p$, though to avoid encumbering
the notation, we will largely suppress the dependence on {\bf k} in the
following.
Because of the relation between $d$ and $\bf k$
the only integral point inside $\Delta$ is $\vone\define (1,1,1,1,1)$.

The Newton polyhedron lives in the four dimensional sublattice of $\IZ^5$
defined by:
$$\Lambda = \lbrace {\bf m} \in \IZ^5 \mid {\bf k} \cdot {\bf m} = d \rbrace$$
or, after translating the origin to the interior point \vone ~and setting
${\bf m}' = {\bf m} - \vone$, by:
$$\Lambda = \lbrace {\bf m}' \in \IZ^5 \mid {\bf k} \cdot {\bf m}' = 0
 \rbrace~.$$
We denote by V the dual lattice to $\Lambda$.
%
The corresponding vector spaces in which these lattices are embedded
are $\Lambda_{\IR} = \Lambda \otimes \IR$ and
$\hbox{V}_{\IR} = \hbox{V} \otimes
\IR$.
Inside these vector spaces the Newton polyhedron and its dual are defined as:
$$\eqalign{\Delta({\bf k}) &= \hbox{ convex hull of} \quad \lbrace {\bf m}' \in
\Lambda({\bf k}) \mid  m'_i \geq -1,  i=1, \ldots ,5 \rbrace \cr
\nabla({\bf k}) &= \lbrace {\bf x} \mid \langle
  {\bf x}, {\bf y} \rangle \geq -1, \forall {\bf y} \in \Delta \rbrace\cr}$$
All the Newton polyhedra constructed this way from $\cal L$, have been checked
 to be reflexive \cite{\rCDK}.
 Reflexivity means (in geometrical terms) that:
 \item{ 1.}$\Delta$ has integer vertices
 \item{ 2.}There is one and only one interior point in  $ \Delta$
 \item{ 3.}The equation of any face of codimension 1, which we write
as $c_1 y_1 + \cdots + c_4 y_4 = 1$ , has coefficients $c_1, \ldots ,
c_4$ that are integers with no common factor.

\noindent The reflexivity condition is important since a polyhedron is
reflexive
if and only if it is the support of global sections of the anticanonical
sheaf on
a Gorenstein Fano variety $\ca{V}_{(\Sigma, \nabla)}$~\cite{\rB}. Any such
section is a linear combination of monomials that correspond to integral
points in
$\Delta$. This variety is a blow up of the $\IP_4^{\bf k}$ that
we started with.
A hypersurface in such a variety, the zero locus of homogeneous
polynomials of fixed degree, admits a \cy\ resolution.
The fan of the embedding variety $\Sigma$ is the fan over a triangulation
of the
faces of the dual polyhedron $\nabla$.

There are two ways of constructing refinements of the fan
$\, \Sigma$. One is to take all rays supported by integral points
$x \in \nabla \cap \hbox{V}$
in the dual polyhedron and, given a triangulation, to construct the
respective fan (we may still be left with cones of volume \Footnote{If all
maximal cones of the fan have volume equal to 1, then the variety is smooth.}
greater than~1, but we are guaranteed that the \cy\ hypersurfaces are going to
be
smooth). In this case we have a crepant \Footnote{A map
is crepant if it preserves the canonical class: in both varieties, \cy\
hypersurfaces can be defined in terms of the same set of functions.}
morphism of toric varieties, $\phi :
\ca{V}_{(\Sigma ', \nabla )} \longrightarrow \ca{V}_{(\Sigma,
\nabla )}$. When we refine a fan we say that we blow up the associated
variety. The inverse procedure will be called a blow down. The variety
$\ca{V}_{(\Sigma ', \nabla )}$ is obtained from $\ca{V}_{(\Sigma, \nabla )}$ by
toric resolutions that do not affect the canonical class: the family of
functions
that may be used to define a
\cy\ hypersurface is not changed in the process. The other way of refining the
fan
is to choose
additional rays that do not correspond to integral points in $
\nabla $. These rays intersect the facets of $
\nabla $ in non-integral points. In some cases the
primitive integer vectors along these rays may define a new polyhedron
$\hat{\nabla}$ which is also reflexive. Even
though the canonical class of the variety has been affected in the
process (there is a new set of functions that we can use in defining \cy\
hypersurfaces) the reflexivity of the polyhedron $\hat{\nabla}$ associated with
the
new variety guarantees the existence of \cy\ hypersurfaces. We claim that
whenever a variety $ \ca{V}_{(\Sigma,
\nabla )}$ admits such a
noncrepant blow up there is an isomorphism between the induced deformation
of  a subset of the \cy\
family originally embedded in $ \ca{V}_{(\Sigma,\nabla )}$ and
the Calabi-Yau family of $\ca{V}_{(\hat{\Sigma} , \hat{\nabla} )}$.

\subsection{Connectivity of Moduli Spaces}
As mentioned in the introduction we aim to show the connectedness of the moduli
spaces for the 7555 families of \cys\ that can be obtained as
transverse hypersurfaces in projective varieties~\cite{{\rKS,\rMaxSkI}}. The
constructive method outlined above tells us that each polyhedron $
\Delta $ lives in the hyperplane defined by the weight vector
$\bf k$. The sublattice $\Lambda $ generated by this hyperplane in
$\IZ^5$ has relative volume $\sqrt{\sum_{i=1}^5 k_i^2}$. Once we find an
appropriate basis for $\Lambda $ we express the coordinates of
all integral points with respect to it. Note that there are an infinite
number of ways to choose a basis for any lattice of dimension greater
then 1 and we will identify polyhedra $ \Delta(\bf k) $ and $
\Delta(\bf k')$ if there is a $GL(4,\IZ)$ bijection between
them. Because the group $GL(4,\IZ)$ is volume preserving we do not change the
structure of the fans supported by $ \Delta$. Each maximal cone
will preserve its volume as well as the number of integral points it
contains. This shows that the two varieties are indeed isomorphic, $
\ca{V}_{(\Sigma,\nabla({\bf k}))} \cong  \ca{V}_{(\Sigma ',\nabla({\bf k
'}))}$, and this is also
true for the respective families of \cy\ hypersurfaces.
When we say that  $ \Delta_2 \subset  \Delta_1$,
we mean that there is a transformation that matches all the
points in  $ \Delta_2$ to a subset of points in $  \Delta_1$.
For example, if the point $a$ in $ \Delta_1$ corresponds to the
monomial ${\bf x}^{\bf m}$ and the point $b$ in $ \Delta_2$
corresponds to the monomial ${\bf y}^{\bf n}$, then identifying $a$
and $b$ implies ${\bf x}^{\bf m} = {\bf y}^{\bf n}$.

Let us now examine more closely the consequences of the inclusion
$\Delta_2 \subset \Delta_1$.\hfil\break
We have:
$$
\matrix{ \ca{V}_{(\Sigma_2,\nabla_2)}      & ~~~~
&\ca{V}_{(\Sigma_1,\nabla_1)}\cropen{3pt}
        \updownarrow   & ~~~~   &\updownarrow\cropen{3pt}
        \kern4pt \Delta_2   &\subset &\kern5pt \Delta_1\cropen{3pt}
        \updownarrow     & ~~~~   &\updownarrow\cropen{3pt}
        \kern4pt \nabla_2 &\supset &\kern5pt \nabla_1\cr}
$$
where $\nabla_2$ is the support of
a noncrepant toric deformation of $ \ca{V}_{(\Sigma_1,\nabla_1)}$ ($\Sigma_2$
is a refinement of $\Sigma_1$), such
that
$$ \ca{V}_{(\Sigma_2,\nabla_2)} \buildrel \phi \over
\longrightarrow  \ca{V}_{(\Sigma_1,\nabla_1)}$$
is a proper birational morphism of toric varieties.

Consider ${\cal M}_2
\subset \ca{V}_{(\Sigma_2,\nabla_2)}$ to be a generic
hypersurface. The points in $\nabla_2 \cap
\hbox{V}_2$ correspond to a subset of monomials
in $\nabla_1 \cap \hbox{V}_1$ that define
a certain hypersurface ${\cal M}_1^\sharp \subset
\ca{V}_{(\Sigma_1,\nabla_1)}$.
The crucial observation is
that the pullback of ${\cal M}_1^\sharp$ to $\ca{V}_{(\Sigma_2,\nabla_2)}$
under $\phi$ is isomorphic to ${\cal M}_2 \,$\cite{{\rCDK}}.
Otherwise said, if $\ca{V}_{(\Sigma_1,\nabla_1)}$ is
blown up to $\ca{V}_{(\Sigma_2,\nabla_2)}$, then the induced
deformation of ${\cal M}_1^\sharp$ is isomorphic to ${\cal M}_2$.

$$
\matrix{
 {\cal M}_2&\leftarrow{\rm birational~ map} \rightarrow &{\cal M}^\sharp_1\cr
\cap& ~~~~                                                   &\cap\cr
\ca{V}_{(\Sigma_2,\nabla_2)}& \rightarrow  {\rm
non~crepant ~blow up} \rightarrow&
\ca{V}_{(\Sigma_1,\nabla_1)} \cr}
$$

What do the inclusions  \eqref{inclusion}, $\Delta_1 \supset \Delta_2 \subset
\Delta_3$, tell us?
There is a special sub-family of hypersurfaces  ${\cal M}_1^\sharp \subset
\ca{V}_{(\Sigma_1,\nabla_1)}$ that can be ``deformed'' into the full family of
hypersurfaces  ${\cal M}_2 \subset \ca{V}_{(\Sigma_2,\nabla_2)}$, that
in turn can be ``deformed'' into a special sub-family  ${\cal M}_3^\sharp
\subset \ca{V}_{(\Sigma_3,\nabla_3)}$.
We remark that the resolution of either  ${\cal M}_1^\sharp$ or
${\cal M}_3^\sharp$
allows us to control all the polynomial deformations of~${\cal M}_2$.
\subsection{Illustration of the Method}
We want to show how the above analysis applies to a pair of varieties.
Consider the manifold
$\IP^{(24,51,133,416,624)}[1248]^{h_{11}=214}_{h_{21}=10}$.
There are 18 points in the
polyhedron and 12 of these are shared with the polyhedron associated
with $\IP^{(54,56,151,522,783)}[1566]^{h_{11}=251}_{h_{21}=5}$.
We are going to look at the
vertices of the interior polyhedron and see how they relate to points of
the exterior polyhedron.
A basis for the sublattice defined by ${\bf k}_1 = (24,51,133,416,624)$
is:
$$\eqalign{
{\bf v}_1 &= (-26,\-0,\-0,\-0,\-1)\cr
{\bf v}_2 &= (-25,\-1,\-1,\-1,\-0) \cr
{\bf v}_3 &= (-52,\-0,\-0,\-3,\-0) \cr
{\bf v}_4 &= (-17,\-8,\-0,\-0,\-0)}$$
The points that correspond to the vertices of the inside polyhedron are
$$\eqalign{
{\bf  a}_1 &= (\-1,-1,\-0,\-0) = (\-0,\-0,\-0,\-0,\-2) \cr
{\bf a}_2 &= (-1,-1,\-1,\-0) = (\-0,\-0,\-0,\-3,\-0) \cr
{\bf a}_3 &= (-1,\-8,-3,-1) = (\-0,\-1,\-9,\-0,\-0) \cr
{\bf a}_4 &= (-1,\-2,-1,\-0) = (\kern3pt 29,\-3,\-3,\-0,\-0) \cr
{\bf a}_5 &= (-1,-1,\-0,\-3) = (\-1,\kern3pt 24,\-0,\-0,\-0)}$$
The vectors on the extreme right with 5 components are the $\bf m$
vectors from which we can read the monomials to which they correspond.
Turning now to the second manifold, we find that
a basis for the sublattice defined by ${\bf k}_2 = (54,56,151,522,783)$
is:
$$\eqalign{{\bf w}_1 &= (\kern3pt 29,\-0,\-0,\-0,-2) \cr {\bf w}_2 &=
(\-1,\-1,\-1,\-1,-1) \cr
{\bf w}_3 &=
(\-0,\-0,\-0,\-3,-2) \cr {\bf w}_4 &= (\kern3pt 10,\-9,\-0,\-1,-2)}$$ The
vertices of the interior
polyhedron are:
$$\eqalign{{\bf b}_1 &= (\-0,-1,\-0,\-0) = (\-0,\-0,\-0,\-0,\-2) \cr
{\bf b}_2 &= (\-0,-1,\-1,\-0) = (\-0,\-0,\-0,\-3,\-0) \cr
{\bf b}_3 &= (\-0,\-9,-3,-1) = (\-0,\-1,\kern3pt 10,\-0,\-0) \cr
{\bf b}_4 &= (-1,-1,-1,\-3) = (\-1,\kern3pt 27,\-0,\-0,\-0) \cr
{\bf b}_5 &= (\-1,-1,\-0,\-0) = (\kern3pt 29,\-0,\-0,\-0,\-0)}$$
\vskip10pt
Define $ {\bf A}_i^j = ({\bf a}_i)^j$ and ${\bf B}_i^j = ({\bf b}_i)^j$. Then
$\bf A =\bf B \bf T$ where
$${\bf T} \, = \, \left(\matrix{-2&0&0&3\cr
                  -1&1&0&0\cr
                  -2&0&1&0\cr
                  -2&1&0&1\cr}\right)$$
If we take the space $\IP^{(24,51,133,416,624)}[1248]^{h_{11}=214}_{h_{21}=10}$
to have
homogeneous coordinates
${x_i}$ and the space $\IP^{(54,56,151,522,783)}[1566]^{h_{11}=251}_{h_{21}=5}$
to have
homogeneous coordinates ${y_i}$ then the correspondence ${\bf a}_i
\leftrightarrow {\bf b}_i$ gives the following identification of
monomials:
$$\eqalign{x_5^2 &= y_5^2\cr
           x_4^2 &= y_4^2\cr
           x_{2}x_3^9 &= y_{2} \, y_3^{10} \cr
           x_{1}^{29} \, x_{2}^{3} \, x_{3}^{3} &= y_{1} \, y_{2}^{27}\cr
           x_{1} \, x_2^{24} &= y_1^{29} \cr}$$
which gives in turn the birational map between the two varieties:
$$\eqalign{y_1 &= x_1^{1/29} \, x_2^{24/29}\cr
           y_2 &= x_1^{840/783} \, x_2^{63/783}x_3^{1/9}\cr
           y_3 &= x_1^{-84/783} \, x_2^{72/783}x_3^{8/9}\cr
           y_4 &= x_4 \cr
           y_5 &= x_5}$$
The map is one to one despite appearances. This must be the case in
virtue of our analysis in \SS 3.2. and may be explicitly checked
by using the scaling properties of the two sets of coordinates.
To conclude, we have a sub-family of hypersurfaces in the variety
described by the exterior polyhedron given by the zero locus of the
polynomial
$$\eqalign{p^{\sharp} =& ~x_1x_2x_3x_4x_5 + x_5^5 + x_4^3 + x_2x_3^9 +
x_1^2x_2^2x_3^2x_4^2 + x_1^{29}x_2^3x_3^3\cr & + x_1^3x_2^3x_3^3x_5 +
x_1^4x_2^4x_3^4x_4 + x_1^6x_2^6x_3^6 + x_1^{10}x_2^9x_3x_4 +
x_1^{12}x_2^{11}x_3^3 + x_1x_2^{24}}$$
that can be blown up to a generic hypersurface in the variety
associated with the interior polyhedron given by the general
polynomial
$$\eqalign{p^{\flat} =& ~y_1y_2y_3y_4y_5 + y_5^2 + y_4^3 + y_2y_3^{10} +
y_1^2y_2^2y_3^2y_4^2 + y_1y_2^{27}\cr & + y_1^3y_2^3y_3^3y_5 +
y_1^4y_2^4y_3^4y_4 + y_1^6y_2^6y_3^6 + y_1^{10}y_2^9y_4 +
y_1^{12}y_2^{11}y_3^2 + y_1^{29}}.$$
\newpage
\section{comp}{The Computation}
In principle, the task of determining all of the reflexive sub-polyhedra
(RSP's) of a given reflexive
polyhedron (RP) is straightforward. Suppose the RP in question has $N$ points.
One can then first look
for all RSP's with $N-1$ points. There are $N-1$ candidates (recall that the
interior point of the RP must
also be the interior point of the RSP).
For each candidate, one would have to determine if it is reflexive. This
requires
identifying the faces, and determining their equations. In general, there are
${N-1}\choose {M-1}$
candidates for RSPs with $M$ points. So, the total number of candidates one
would have to consider is:
$$\sum_{M=6}^{N-1} {{N-1}\choose {M-1}} \approx 2^{N-1}.$$ Since many RP's have
$N>200$, this is not feasible.
While this  method is complete, and would have to be followed in
order to establish that
a group of RP's are {\it not} directly connected, it is thankfully not
necessary if one is trying
to establish their connectedness. One might imagine that it might happen that
all of the
RP's under consideration contain a particular RSP, and connectedness would
follow immediately,
irrespective of any other RSP's that they might contain. This turns out not to
be the case, but is the spirit in which we have attacked the
problem.

There is no such RSP. To see this is simple. There do exist reflexive polyhedra
with 6 points. These are simplices in four dimensions (recall
that there is always one interior point), and they cannot contain any RSP's,
hence they themselves would
have to be the magical RSP's. However, there are three inequivalent RP's with 6
points. Although the simplest guess fails, nevertheless one might hope to
restrict one's search for RSP's to some
simple objects, such as simplices.

Our initial strategy was to identify the reflexive simplices within each of the
7555
RP's. In fact we should identify those reflexive simplices which
themselves contain no
reflexive simplices. We will refer to such objects as 5--vertex irreducible
simplices, that is they are simplices that contain no 5--vertex RSP's. In
general, any
reflexive polyhedron which does not contain any $n$-vertex reflexive polyhedra
(apart from itself) is an $n$-vertex irreducible polyhedron.
The combinatoric barrier is
now reduced from $2^{N-1}$ to ${N-1}\choose 5$, which is a good deal more
manageable.
In fact, it was a relatively quick matter to decompose all RP's with
$N\le 120$ into 5-vertex irreducible simplices. Very early in this procedure
(i.e. after decomposing a
small fraction of these RP's), a set of 41 5-vertex irreducible simplices was
generated,
ranging in size from $N=6$ to $N=26$ points. It turned
out that these were the only ones that were generated by this procedure, though
we have no proof that
others do not exist. Among them are 18 that are not on the list of 7555 that we
started
 from. Being simplices, these must correspond to Fermat polynomials, but since
they are not all in the list, they must in fact correspond to manifolds of the
form ${\lbrace p=0 \rbrace}/G$, where $p$ is Fermat and $G$ is a group of
automorphisms.

Establishing the connectedness of all of those RP's that contain one or more of
these simplices
is then a matter of showing that these 41 simplices are connected to each
other. Let us denote the set of 5-vertex irreducible simplices
contained in the $i$th RP as ${\bf V}^5_i$. We define ${\bf C}^5_1 ={\bf
V}^5_1$.
 In general:
$${\bf C}^5_i = {\bf C}^5_{i-1} \cup \delta_{i-1,i}~,\quad \hbox{where}
\quad \delta_{i,j} = \cases{\emptyset &if ${\bf C}_i^5\cap {\bf V}_j^5
=\emptyset$\cr
                            {\bf V}_j^5     &if ${\bf C}_i^5\cap {\bf
V}_j^5\ne\emptyset$\cr}
 $$
If for any $i$, ${\bf C}^5_i$ contains all 41 of the 5-vertex irreducible
simplices, then their
connectedness has been established. This procedure, which is a sufficient but
not necessary condition, can be described as follows.
For each polyhedra with $N\le 120$, we have a list of which of the 41 reflexive
simplices that it contains, ${\bf V}^5_i$. Clearly, all of the simplices that
are contained in ${\bf V}^5_1$ are connected, since they are all connected to
the first polyhedron. So this is our initial list of connected simplices.
Now, successively examine each of the ${\bf V}^5_i$. Whenever there is a
simplex in ${\bf V}^5_i$ that is also in our list of connected simplices, we
add {\it all} of the simplices in ${\bf V}^5_i$ to our list. As soon as this
list contains all 41 simplices, we have shown that they are connected. This is
indeed the case, hence all those
RP's with
$N\le 120$ and ${\bf V}^5_i \neq \emptyset$, which amounts to 6133 RP's, are
connected.

For those RP's with $N>120$, we take advantage of the apparent completeness of
the list of 5-vertex
irreducible simplices. So instead of finding {\it every} 5-vertex irreducible
simplex in these
larger objects, we stop our search as soon as we find {\it one}. This approach
succeeds, in that all
of these larger RP's  contain at least one of the 41  previously found 5-vertex
irreducible
simplices. This accounts for a further 1280 of the RP's.

We were left with 142 RP's that do not contain any reflexive simplices.
As they are relatively small ($N\le 28$), it was feasible to try and search for
RSPs that had
6 vertices, and were themselves 6-vertex irreducible. Applying this procedure,
we found that
132 of the remainder contained 6-vertex irreducible 6-vertex RSPs, and
 that 10 did not.

In order to see if these 132 RP's were connected to the 6133 RP's that were
already known
to be interconnected, it was necessary to make sure that all of the 6-vertex
irreducible 6-vertex
RSPs that were generated by the former list, were also generated by some subset
(hopefully
a small one) of the latter list. In practice, it was easy to establish this for
123 of
them. The remaining 9 are difficult cases.

We then attempted to decompose these 9, plus the 10
RP's that contained neither reflexive simplices, nor reflexive 6-vertex
polyhedra,
into 7-vertex irreducible 7-vertex RSPs. Seventeen of them yielded to this
procedure, and were
easily connected to the other 7536 RP's. This leaves us with two polyhedra.
They correspond
to $\IP^{(21,24,82,111,119)}_4[357]$ and $\IP^{(18,21,58,77,78)}_4[252]$. They
each contain the same
6-vertex irreducible 6-vertex RSP -- one which we could not easily locate in
another RP.
Let us call it $\Pi$.

In order to take care of these last two polyhedra, we make use of the
elementary observation made previously, that if $\D_2$ and $\D_1$ are reflexive
polyhedra, and  $\D_2\subset \D_1$, then $\nabla_1 \subset \nabla_2$, where
$\nabla_2$ and $\nabla_1$ are the dual polyhedra of $\D_2$ and $\D_1$.
First, we established that the dual of $\Pi$ was one of the RP's
from our original list. It corresponds to $\IP^{(1,1,4,6,6)}_4[18]$. Thus the
duals of whatever RSPs are contained within this RP, contain $\Pi$. In fact,
the RP corresponding
to $\IP^{(1,1,4,6,6)}_4[18]$ contains
a simplex whose dual polyhedron corresponds to $\IP^{(1,1,1,6,9)}_4[18]$. So,
we learn that
$\IP^{(21,24,82,111,119)}_4[357]$ and $\IP^{(18,21,58,77,78)}_4[252]$
each contain an RSP that is also contained in the RP corresponding to
$\IP^{(1,1,1,6,9)}_4[18]$.
Furthermore, the RP corresponding to $\IP^{(1,1,1,6,9)}_4[18]$ is one of the
RP's that had previously been
connected via 5-vertex irreducible simplices.

We have thus succeeded in establishing that the moduli spaces of all three
dimensional CICYs and all
transverse polynomials in four dimensional weighted projective spaces are
interconnected.
\vskip0.7in
\acknowledgements
We gratefully acknowledge fruitful conversations with S.~Katz, S.~Keel,
D.~Morrison and D.~Saltman. This work was supported by NSF Grants PHY95-11632
and PHY94-07194 and the Robert A.~Welch Foundation. P.~Candelas and
M.~Mandelberg would like to thank SBITP for
hospitality while part of this work was being done.
\vskip0.7in
\chapno=-1
\section{app}{Appendix: The $n$-Vertex Irreducible RP's that Connect the
Manifolds}
Here we give some information regarding those polyhedra which
we have used to connect the moduli spaces of all the manifolds of the list
\ca{L}.
For each polyhedron, the following information is given: the number of points
it
contains, its associated hodge numbers, and, in the case that it corresponds to
one
of the polyhedra corresponding to one or more of the spaces in
$\cal L$, the weights of those spaces, enclosed in square brackets. Instead, if
it
does not correspond to any entry on
$\cal L$, the vertices of the polyhedron are given, enclosed in parentheses.
There
are three tables, one for each $n$-vertex irreducible $n$-vertex RP's with
$n=5,6$
and $7$.
\newpage
\centerline{{\bf Table 1:} 5-vertex irreducible 5-vertex reflexive polyhedra}
\vskip10pt
\line{\vbox{\offinterlineskip\halign{\strut#&\vrule#&
\hfil\quad\sixrm#\quad&#\vrule&
\hfil\quad\sixrm#\quad&#\vrule&
\hfil\quad\sixrm#\quad&#\vrule&
\quad\sixrm#\quad\hfil&#\vrule\cr
\noalign{\hrule}
&
  &\hbox{\tenrm Pts}
    &&$h_{11}$\hfil
      &&$h_{21}$\hfil
        &&\hbox{\eightrm Vertices/Weights}&\cr
\noalign{\hrule\vskip3pt\hrule}
&   &6   &&103   &&1   && [52,60,63,75,125] &\cropen{-3pt}
&   &    &&      &&    && [48,50,60,63,79] &\cr
\noalign{\hrule}
&   &6   &&101   &&1   && [41,48,51,52,64] &\cropen{-3pt}
&   &    &&      &&    && [51,60,64,65,80] &\cr
\noalign{\hrule}
&   &6    &&21   &&1   && (0,1,-1,0) &\cropen{-3pt}
&   &     &&     &&    && (-2,1,1,-1) &\cropen{-3pt}
&   &     &&     &&    && (-1,-1,0,2) &\cropen{-3pt}
&   &     &&     &&    && (0,0,-1,0) &\cropen{-3pt}
&   &     &&     &&    && (3,-1,1,-1) &\cr
\noalign{\hrule}
&   &7    &&149  &&1   && [75,84,86,98,343] &\cropen{-3pt}
&   &     &&     &&    && [43,48,56,98,147] &\cropen{-3pt}
&   &     &&     &&    && [48,49,56,86,153] &\cropen{-3pt}
&   &     &&     &&    && [42,43,49,75,134] &\cr
\noalign{\hrule}
&   &7    &&145  &&1   && [73,80,90,162,405] &\cropen{-3pt}
&   &     &&     &&    && [64,72,73,115,324] &\cropen{-3pt}
&   &     &&     &&    && [40,45,73,81,166] &\cropen{-3pt}
&   &     &&     &&    && [40,45,63,64,148] &\cr
\noalign{\hrule}
&   &7    &&112  &&4   && [42,48,57,109,128] &\cr
\noalign{\hrule}
&   &7    &&86   &&2   && [27,36,49,56,84] &\cropen{-3pt}
&   &     &&     &&    && [27,36,41,52,60] &\cr
\noalign{\hrule}
&   &7    &&89   &&5   && [36,40,89,99,132] &\cr
\noalign{\hrule}
&   &7    &&38   &&2   && (1,-1,1,-1) &\cropen{-3pt}
&   &     &&     &&    && (-2,-1,1,1) &\cropen{-3pt}
&   &     &&     &&    && (0,-1,-1,2) &\cropen{-3pt}
&   &     &&     &&    && (-1,-1,0,0) &\cropen{-3pt}
&   &     &&     &&    && (0,2,-1,0) &\cr
\noalign{\hrule}
&   &8    &&165  &&3   && [60,66,74,163,363] &\cropen{-3pt}
&   &     &&     &&    && [40,44,74,121,205] &\cropen{-3pt}
&   &     &&     &&    && [36,40,66,89,165] &\cr
\noalign{\hrule}
&   &8    &&128  &&2   && [25,30,66,88,121] &\cropen{-3pt}
&   &     &&     &&    && [25,30,54,82,109] &\cropen{-3pt}
&   &     &&     &&    && [30,41,50,54,125] &\cr
\noalign{\hrule}
&   &8    &&103  &&7   && [30,39,84,127,140] &\cr
\noalign{\hrule}
&   &8    &&101  &&5   && (-1,0,0,1) &\cr
&   &     &&     &&    && (-1,0,1,0) &\cropen{-3pt}
&   &     &&     &&    && (-1,1,0,0) &\cropen{-3pt}
&   &     &&     &&    && (-1,0,0,0) &\cropen{-3pt}
&   &     &&     &&    && (11,-1,-4,-2) &\cr
\noalign{\hrule}
&   &8    &&60   &&6   && (0,-1,0,1) &\cropen{-3pt}
&   &     &&     &&    && (2,-1,0,0) &\cropen{-3pt}
&   &     &&     &&    && (0,-1,1,0) &\cropen{-3pt}
&   &     &&     &&    && (0,-1,0,0) &\cropen{-3pt}
&   &     &&     &&    && (-3,5,-1,-1) &\cr
\noalign{\hrule}}}
\hfill
\vbox{\offinterlineskip\halign{\strut#&\vrule#&
\hfil\quad\sixrm#\quad&#\vrule&
\hfil\quad\sixrm#\quad&#\vrule&
\hfil\quad\sixrm#\quad&#\vrule&
\quad\sixrm#\quad\hfil&#\vrule\cr
\noalign{\hrule}
&
  &\hbox{\tenrm Pts}
    &&$h_{11}$\hfil
      &&$h_{21}$\hfil
        &&\hbox{\eightrm Vertices/Weights}&\cr
\noalign{\hrule\vskip3pt\hrule}
&   &8    &&63   &&3   && (0,-1,1,0) &\cropen{-3pt}
&   &     &&     &&    && (-2,5,-1,-2) &\cropen{-3pt}
&   &     &&     &&    && (0,-1,0,1) &\cropen{-3pt}
&   &     &&     &&    && (0,-1,0,0) &\cropen{-3pt}
&   &     &&     &&    && (2,-1,0,1) &\cr
\noalign{\hrule}
&   &8    &&31   &&13  && (-3,2,0,1) &\cropen{-3pt}
&   &     &&     &&    && (0,-1,3,-1) &\cropen{-3pt}
&   &     &&     &&    && (-1,1,-1,1) &\cropen{-3pt}
&   &     &&     &&    && (0,-1,0,0) &\cropen{-3pt}
&   &     &&     &&    && (4,0,-2,-1) &\cr
\noalign{\hrule}
&   &9    &&148  &&4   && [24,28,83,90,135] &\cropen{-2pt}
&   &     &&     &&    && [24,28,77,78,129] &\cropen{-2pt}
&   &     &&     &&    && [24,42,52,71,147] &\cr
\noalign{\hrule}
&   &9    &&83   &&3   && (0,-1,0,0) &\cropen{-3pt}
&   &     &&     &&    && (0,-1,1,0) &\cropen{-3pt}
&   &     &&     &&    && (2,-1,0,0) &\cropen{-3pt}
&   &     &&     &&    && (-1,-1,0,1) &\cropen{-3pt}
&   &     &&     &&    && (-1,7,-4,-1) &\cr
\noalign{\hrule}
&   &9    &&75   &&3   && (2,1,-1,-1) &\cropen{-3pt}
&   &     &&     &&    && (-1,0,0,0) &\cropen{-3pt}
&   &     &&     &&    && (-1,0,0,1) &\cropen{-3pt}
&   &     &&     &&    && (0,-1,-1,2) &\cropen{-3pt}
&   &     &&     &&    && (1,-2,6,-1) &\cr
\noalign{\hrule}
&   &9    &&55   &&7   && (0,-1,1,0) &\cropen{-3pt}
&   &     &&     &&    && (-1,1,-1,1) &\cropen{-3pt}
&   &     &&     &&    && (-2,3,1,-2) &\cropen{-3pt}
&   &     &&     &&    && (0,-1,0,0) &\cropen{-3pt}
&   &     &&     &&    && (2,-1,0,0) &\cr
\noalign{\hrule}
&   &9    &&43   &&3   && (3,0,0,-1) &\cropen{-3pt}
&   &     &&     &&    && (1,-1,1,0) &\cropen{-3pt}
&   &     &&     &&    && (1,0,2,-1) &\cropen{-3pt}
&   &     &&     &&    && (0,-1,0,0) &\cropen{-3pt}
&   &     &&     &&    && (-5,5,-3,2) &\cr
\noalign{\hrule}
&   &9    &&29   &&5   && (-1,0,-1,2) &\cropen{-3pt}
&   &     &&     &&    && (-1,1,-1,0) &\cropen{-3pt}
&   &     &&     &&    && (-1,0,-1,0) &\cropen{-3pt}
&   &     &&     &&    && (3,-1,1,-1) &\cropen{-3pt}
&   &     &&     &&    && (-1,1,3,-2) &\cr
\noalign{\hrule}
&   &10   &&272  &&2   && [91,96,102,578,867] &\cropen{-2.2pt}
&   &     &&     &&    && [64,68,91,355,578] &\cropen{-2.2pt}
&   &     &&     &&    && [48,51,91,289,388] &\cropen{-2.2pt}
&   &     &&     &&    && [36,51,64,187,274] &\cr
\noalign{\hrule}
&   &10   &&143  &&7   && [40,45,143,152,380] &\cr
\noalign{\hrule}
&   &10   &&105  &&3   && [33,36,40,89,198] &\cr
\noalign{\hrule}}}}
\newpage
\centerline{{\bf Table 1(cont'd):} 5-vertex irreducible 5-vertex reflexive
polyhedra}
\vskip10pt
\line{
\vbox{\offinterlineskip\halign{\strut#&\vrule#&
\hfil\quad\sixrm#\quad&#\vrule&
\hfil\quad\sixrm#\quad&#\vrule&
\hfil\quad\sixrm#\quad&#\vrule&
\quad\sixrm#\quad\hfil&#\vrule\cr
\noalign{\hrule}
&
  &\hbox{\tenrm Pts}
    &&$h_{11}$\hfil
      &&$h_{21}$\hfil
        &&\hbox{\eightrm Vertices/Weights}&\cr
\noalign{\hrule\vskip3pt\hrule}
&   &10   &&21   &&9   && (2,-1,0,0) &\cropen{-1.9pt}
&   &     &&     &&    && (-1,-1,-1,2) &\cropen{-1.9pt}
&   &     &&     &&    && (-1,-1,1,0) &\cropen{-1.9pt}
&   &     &&     &&    && (-1,-1,1,1) &\cropen{-1.9pt}
&   &     &&     &&    && (-1,5,-1,-3) &\cr
\noalign{\hrule}
&   &10   &&17   &&21  && (2,-1,-1,0) &\cropen{-1.9pt}
&   &     &&     &&    && (0,0,-1,0) &\cropen{-1.9pt}
&   &     &&     &&    && (-1,3,-1,0) &\cropen{-1.9pt}
&   &     &&     &&    && (-1,-1,-1,1) &\cropen{-1.9pt}
&   &     &&     &&    && (0,-1,4,-1) &\cr
\noalign{\hrule}
&   &11   &&164  &&8   && (1,-1,0,0) &\cropen{-1.9pt}
&   &     &&     &&    && (0,-1,1,0) &\cropen{-1.9pt}
&   &     &&     &&    && (0,-1,0,0) &\cropen{-1.9pt}
&   &     &&     &&    && (-1,5,-3,-1) &\cropen{-1.9pt}
&   &     &&     &&    && (-2,-1,0,4) &\cr
\noalign{\hrule}
&   &11   &&131  &&11  && [24,33,138,173,184] &\cr
\noalign{\hrule}
&   &11   &&69   &&9   && (-1,0,0,1) &\cropen{-1.9pt}
&   &     &&     &&    && (-1,1,0,0) &\cropen{-1.9pt}
&   &     &&     &&    && (-1,-2,4,0) &\cropen{-1.9pt}
&   &     &&     &&    && (-1,0,0,0) &\cropen{-1.9pt}
&   &     &&     &&    && (5,-2,-2,-1) &\cr
\noalign{\hrule}
&   &11   &&59   &&11  && (-1,0,1,0) &\cropen{-1.9pt}
&   &     &&     &&    && (-1,2,-1,0) &\cropen{-1.9pt}
&   &     &&     &&    && (0,-1,0,1) &\cropen{-1.9pt}
&   &     &&     &&    && (0,-1,0,0) &\cropen{-1.9pt}
&   &     &&     &&    && (6,-1,0,-1) &\cr
\noalign{\hrule}}}
\hfill
\vbox{\offinterlineskip\halign{\strut#&\vrule#&
\hfil\quad\sixrm#\quad&#\vrule&
\hfil\quad\sixrm#\quad&#\vrule&
\hfil\quad\sixrm#\quad&#\vrule&
\quad\sixrm#\quad\hfil&#\vrule\cr
\noalign{\hrule}
&
  &\hbox{\tenrm Pts}
    &&$h_{11}$\hfil
      &&$h_{21}$\hfil
        &&\hbox{\eightrm Vertices/Weights}&\cr
\noalign{\hrule\vskip3pt\hrule}
&   &11   &&243  &&3   && [24,33,92,173,230] &\cropen{-3pt}
&   &     &&     &&    && [24,44,69,161,254] &\cropen{-3pt}
&   &     &&     &&    && [24,44,63,155,242] &\cr
\noalign{\hrule}
&   &12   &&251  &&5   && [54,56,151,522,783] &\cropen{-3pt}
&   &     &&     &&    && [30,56,87,290,407] &\cr
\noalign{\hrule}
&   &13   &&82   &&10  && (2,-1,1,-1) &\cropen{-3pt}
&   &     &&     &&    && (-1,-1,1,1) &\cropen{-3pt}
&   &     &&     &&    && (-1,1,0,0) &\cropen{-3pt}
&   &     &&     &&    && (0,-1,0,0) &\cropen{-3pt}
&   &     &&     &&    && (5,-1,-5,3) &\cr
\noalign{\hrule}
&   &13   &&35   &&19  && (1,-1,0,0) &\cropen{-3pt}
&   &     &&     &&    && (-1,-1,1,0) &\cropen{-3pt}
&   &     &&     &&    && (-1,-1,3,0) &\cropen{-3pt}
&   &     &&     &&    && (-1,-1,0,1) &\cropen{-3pt}
&   &     &&     &&    && (-1,7,-4,-1) &\cr
\noalign{\hrule}
&   &14   &&271  &&7   && [48,51,181,560,840] &\cropen{-3pt}
&   &     &&     &&    && [36,51,140,403,630] &\cr
\noalign{\hrule}
&   &15   &&227  &&11  && [30,38,234,283,585] &\cropen{-3pt}
&   &     &&     &&    && [20,38,156,195,371] &\cr
\noalign{\hrule}
&   &15   &&103  &&7   && [18,20,57,85,180] &\cr
\noalign{\hrule}
&   &17   &&321  &&9   && [42,46,241,658,987] &\cropen{-3pt}
&   &     &&     &&    && [24,46,141,376,541] &\cr
\noalign{\hrule}
&   &21   &&131  &&11  && [15,16,93,116,240] &\cr
\noalign{\hrule}
&   &26   &&491  &&11  && [41,42,498,1162,1743] &\cropen{-3pt}
&   &     &&     &&    && [36,41,421,996,1494] &\cropen{-3pt}
&   &     &&     &&    && [28,41,332,761,1162] &\cropen{-3pt}
&   &     &&     &&    && [21,41,249,581,851] &\cr
\noalign{\hrule}}}}
\newpage
\centerline{{\bf Table 2:} 6-vertex irreducible 6-vertex reflexive polyhedra}
\vskip10pt
\line{
\vbox{\offinterlineskip\halign{\strut#&\vrule#&
\hfil\quad\sixrm#\quad&#\vrule&
\hfil\quad\sixrm#\quad&#\vrule&
\hfil\quad\sixrm#\quad&#\vrule&
\quad\sixrm#\quad\hfil&#\vrule\cr
\noalign{\hrule}
&
  &\hbox{\tenrm Pts}
    &&$h_{11}$\hfil
      &&$h_{21}$\hfil
        &&\hbox{\eightrm Vertices/Weights}&\cr
\noalign{\hrule\vskip3pt\hrule}
&   &7    &&95   &&2   && [32,42,45,91,105] &\cropen{-3pt}
&   &     &&     &&    && [26,33,48,75,91] &\cropen{-3pt}
&   &     &&     &&    && [26,33,45,60,64] &\cropen{-3pt}
&   &     &&     &&    && [23,28,34,53,55] &\cr
\noalign{\hrule}
&   &7    &&86   &&2   && [27,29,64,72,96] &\cropen{-3pt}
&   &     &&     &&    && [22,29,49,50,75] &\cropen{-3pt}
&   &     &&     &&    && [29,36,75,84,112] &\cropen{-3pt}
&   &     &&     &&    && [27,28,59,63,75] &\cropen{-3pt}
&   &     &&     &&    && [21,29,53,56,65] &\cropen{-3pt}
&   &     &&     &&    && [24,29,50,56,65] &\cr
\noalign{\hrule}
&   &8    &&132  &&2   && [44,48,63,131,286] &\cropen{-3pt}
&   &     &&     &&    && [28,39,48,91,158] &\cr
\noalign{\hrule}
&   &8    &&105  &&3   && (0,-1,0,0) &\cropen{-3pt}
&   &     &&     &&    && (-1,1,0,0) &\cropen{-3pt}
&   &     &&     &&    && (5,1,-1,-2) &\cropen{-3pt}
&   &     &&     &&    && (0,-1,1,0) &\cropen{-3pt}
&   &     &&     &&    && (2,-1,0,0) &\cropen{-3pt}
&   &     &&     &&    && (0,-1,0,1) &\cr
\noalign{\hrule}
&   &8    &&87   &&3   && [22,30,63,95,105] &\cr
\noalign{\hrule}
&   &8    &&43   &&7   && (-1,-1,0,0) &\cropen{-3pt}
&   &     &&     &&    && (0,-1,1,0) &\cropen{-3pt}
&   &     &&     &&    && (1,-1,-1,0) &\cropen{-3pt}
&   &     &&     &&    && (-1,-1,0,1) &\cropen{-3pt}
&   &     &&     &&    && (0,1,0,0) &\cropen{-3pt}
&   &     &&     &&    && (1,4,-1,-1) &\cr
\noalign{\hrule}
&   &8    &&59   &&3   && (0,-1,1,0) &\cropen{-3pt}
&   &     &&     &&    && (1,-1,-1,0) &\cropen{-3pt}
&   &     &&     &&    && (-1,-1,0,1) &\cropen{-3pt}
&   &     &&     &&    && (-1,-1,1,0) &\cropen{-3pt}
&   &     &&     &&    && (0,1,0,0) &\cropen{-3pt}
&   &     &&     &&    && (1,4,-1,-1) &\cr
\noalign{\hrule}
&   &10   &&152  &&6   && [34,36,122,131,323] &\cropen{-3pt}
&   &     &&     &&    && [20,34,64,69,153] &\cr
\noalign{\hrule}
&   &10   &&131  &&3   && [33,40,122,130,325] &\cropen{-3pt}
&   &     &&     &&    && [20,33,61,65,146] &\cr
\noalign{\hrule}
&   &10   &&139  &&5   && (-1,1,0,0) &\cropen{-3pt}
&   &     &&     &&    && (5,0,-1,-1) &\cropen{-3pt}
&   &     &&     &&    && (-1,0,1,0) &\cropen{-3pt}
&   &     &&     &&    && (-1,0,0,0) &\cropen{-3pt}
&   &     &&     &&    && (1,0,-1,1) &\cropen{-3pt}
&   &     &&     &&    && (-1,-1,-1,3) &\cr
\noalign{\hrule}
&   &10   &&111  &&9   && (-1,1,0,0) &\cropen{-3pt}
&   &     &&     &&    && (-1,0,1,0) &\cropen{-3pt}
&   &     &&     &&    && (-1,0,0,1) &\cropen{-3pt}
&   &     &&     &&    && (-1,0,0,0) &\cropen{-3pt}
&   &     &&     &&    && (3,-1,-1,-1) &\cropen{-3pt}
&   &     &&     &&    && (17,-6,-6,-4) &\cr
\noalign{\hrule}}}
\hfill
\vbox{\offinterlineskip\halign{\strut#&\vrule#&
\hfil\quad\sixrm#\quad&#\vrule&
\hfil\quad\sixrm#\quad&#\vrule&
\hfil\quad\sixrm#\quad&#\vrule&
\quad\sixrm#\quad\hfil&#\vrule\cr
\noalign{\hrule}
&
  &\hbox{\tenrm Pts}
    &&$h_{11}$\hfil
      &&$h_{21}$\hfil
        &&\hbox{\eightrm Vertices/Weights}&\cr
\noalign{\hrule\vskip3pt\hrule}
&   &11   &&114  &&6   && (-1,0,1,0) &\cropen{-2pt}
&   &     &&     &&    && (2,-1,1,-1) &\cropen{-2pt}
&   &     &&     &&    && (-1,1,0,0) &\cropen{-2pt}
&   &     &&     &&    && (0,-1,0,0) &\cropen{-2pt}
&   &     &&     &&    && (2,-1,-2,2) &\cropen{-2pt}
&   &     &&     &&    && (3,-1,-3,2) &\cr
\noalign{\hrule}
&   &11   &&101  &&5   && [17,24,99,124,132] &\cropen{-2pt}
&   &     &&     &&    && [17,18,70,93,99] &\cr
\noalign{\hrule}
&   &11   &&70   &&4   && (0,-1,0,0) &\cropen{-3pt}
&   &     &&     &&    && (0,-1,0,1) &\cropen{-3pt}
&   &     &&     &&    && (1,0,0,0) &\cropen{-3pt}
&   &     &&     &&    && (-1,0,2,0) &\cropen{-3pt}
&   &     &&     &&    && (0,3,-2,0) &\cropen{-3pt}
&   &     &&     &&    && (0,6,-3,-1) &\cr
\noalign{\hrule}
&   &13   &&43   &&11  && (-1,-1,0,0) &\cropen{-3pt}
&   &     &&     &&    && (3,-1,-2,0) &\cropen{-3pt}
&   &     &&     &&    && (3,6,-2,-1) &\cropen{-3pt}
&   &     &&     &&    && (-1,-1,1,0) &\cropen{-3pt}
&   &     &&     &&    && (1,3,-1,0) &\cropen{-3pt}
&   &     &&     &&    && (-1,-1,0,1) &\cr
\noalign{\hrule}
&   &14   &&194  &&10  && (1,-1,0,0) &\cropen{-3pt}
&   &     &&     &&    && (0,-1,1,0) &\cropen{-3pt}
&   &     &&     &&    && (0,-1,0,1) &\cropen{-3pt}
&   &     &&     &&    && (0,-1,0,0) &\cropen{-3pt}
&   &     &&     &&    && (-5,9,-2,-2) &\cropen{-3pt}
&   &     &&     &&    && (-16,31,-8,-6) &\cr
\noalign{\hrule}
&   &15   &&183  &&7   && [23,30,182,220,455] &\cropen{-2.5pt}
&   &     &&     &&    && [23,24,141,176,364] &\cropen{-2.5pt}
&   &     &&     &&    && [15,23,91,110,216] &\cr
\noalign{\hrule}
&   &15   &&157  &&9   && (0,-1,0,0) &\cropen{-3pt}
&   &     &&     &&    && (1,-1,0,0) &\cropen{-3pt}
&   &     &&     &&    && (-1,-1,1,0) &\cropen{-3pt}
&   &     &&     &&    && (-1,-1,0,1) &\cropen{-3pt}
&   &     &&     &&    && (-1,19,-5,-4) &\cropen{-3pt}
&   &     &&     &&    && (-1,24,-6,-5) &\cr
\noalign{\hrule}
&   &15   &&166  &&8   && (1,-1,0,0) &\cropen{-3pt}
&   &     &&     &&    && (-1,-1,1,0) &\cropen{-3pt}
&   &     &&     &&    && (-1,-1,0,1) &\cropen{-3pt}
&   &     &&     &&    && (-1,0,0,0) &\cropen{-3pt}
&   &     &&     &&    && (-1,19,-5,-4) &\cropen{-3pt}
&   &     &&     &&    && (-1,24,-6,-5) &\cr
\noalign{\hrule}
&   &15   &&140  &&10  && (-1,2,0,-1) &\cropen{-3pt}
&   &     &&     &&    && (0,0,1,-1) &\cropen{-3pt}
&   &     &&     &&    && (0,0,0,-1) &\cropen{-3pt}
&   &     &&     &&    && (0,-1,0,1) &\cropen{-3pt}
&   &     &&     &&    && (5,0,0,-1) &\cropen{-3pt}
&   &     &&     &&    && (5,0,-1,0) &\cr
\noalign{\hrule}}}}
\newpage
\centerline{{\bf Table 2(cont'd):} 6-vertex irreducible 6-vertex reflexive
polyhedra}
\bigskip
\centerline{
\vbox{\offinterlineskip\halign{\strut#&\vrule#&
\hfil\quad\sixrm#\quad&#\vrule&
\hfil\quad\sixrm#\quad&#\vrule&
\hfil\quad\sixrm#\quad&#\vrule&
\quad\sixrm#\quad\hfil&#\vrule\cr
\noalign{\hrule}
&
  &\hbox{\tenrm Pts}
    &&$h_{11}$\hfil
      &&$h_{21}$\hfil
        &&\hbox{\eightrm Vertices/Weights}&\cr
\noalign{\hrule\vskip3pt\hrule}
&   &15   &&149  &&9   && (-1,2,0,-1) &\cropen{-3pt}
&   &     &&     &&    && (0,0,1,-1) &\cropen{-3pt}
&   &     &&     &&    && (0,0,0,-1) &\cropen{-3pt}
&   &     &&     &&    && (0,-1,0,1) &\cropen{-3pt}
&   &     &&     &&    && (5,0,0,-1) &\cropen{-3pt}
&   &     &&     &&    && (5,1,-1,-1) &\cr
\noalign{\hrule}
&   &15   &&131  &&11  && (-1,2,0,-1) &\cropen{-3pt}
&   &     &&     &&    && (0,0,0,-1) &\cropen{-3pt}
&   &     &&     &&    && (0,-1,0,1) &\cropen{-3pt}
&   &     &&     &&    && (0,-1,1,0) &\cropen{-3pt}
&   &     &&     &&    && (5,0,0,-1) &\cropen{-3pt}
&   &     &&     &&    && (5,0,-1,0) &\cr
\noalign{\hrule}
&   &17   &&295  &&7   && [37,42,216,590,885] &\cropen{-3pt}
&   &     &&     &&    && [28,37,144,381,590] &\cropen{-3pt}
&   &     &&     &&    && [21,37,108,295,424] &\cr
\noalign{\hrule}
&   &22   &&376  &&10  && (-1,0,1,0) &\cropen{-3pt}
&   &     &&     &&    && (-1,0,0,1) &\cropen{-3pt}
&   &     &&     &&    && (-1,1,0,0) &\cropen{-3pt}
&   &     &&     &&    && (-1,0,0,0) &\cropen{-3pt}
&   &     &&     &&    && (13,0,-4,-2) &\cropen{-3pt}
&   &     &&     &&    && (59,-1,-20,-8) &\cr
\noalign{\hrule}}}}
\bigskip
\centerline{{\bf Table 3:} 7-vertex irreducible 7-vertex reflexive polyhedra}
\bigskip
\centerline{
\vbox{\offinterlineskip\halign{\strut#&\vrule#&
\hfil\quad\sixrm#\quad&#\vrule&
\hfil\quad\sixrm#\quad&#\vrule&
\hfil\quad\sixrm#\quad&#\vrule&
\quad\sixrm#\quad\hfil&#\vrule\cr
\noalign{\hrule}
&
  &\hbox{\tenrm Pts}
    &&$h_{11}$\hfil
      &&$h_{21}$\hfil
        &&\hbox{\eightrm Vertices/Weights}&\cr
\noalign{\hrule\vskip3pt\hrule}
&   &8    &&51   &&3   && [20,21,25,26,33] &\cr
\noalign{\hrule}
&   &10   &&92   &&5   && (0,-1,1,0) &\cropen{-3pt}
&   &     &&     &&    && (0,-1,0,1) &\cropen{-3pt}
&   &     &&     &&    && (-1,2,0,-1) &\cropen{-3pt}
&   &     &&     &&    && (0,-1,0,0) &\cropen{-3pt}
&   &     &&     &&    && (1,0,0,0) &\cropen{-3pt}
&   &     &&     &&    && (3,-1,0,0) &\cropen{-3pt}
&   &     &&     &&    && (4,0,-1,0) &\cr
\noalign{\hrule}
&   &13   &&152  &&6   && (1,-1,0,0) &\cropen{-3pt}
&   &     &&     &&    && (-1,-1,1,0) &\cropen{-3pt}
&   &     &&     &&    && (-1,-1,0,1) &\cropen{-3pt}
&   &     &&     &&    && (-1,0,0,0) &\cropen{-3pt}
&   &     &&     &&    && (-1,9,-2,-2) &\cropen{-3pt}
&   &     &&     &&    && (-1,11,-2,-3) &\cropen{-3pt}
&   &     &&     &&    && (-1,20,-4,-5) &\cr
\noalign{\hrule}
&   &14   &&167  &&7   && (-1,1,0,0) &\cropen{-3pt}
&   &     &&     &&    && (-1,0,0,1) &\cropen{-3pt}
&   &     &&     &&    && (4,0,-1,-1) &\cropen{-3pt}
&   &     &&     &&    && (-1,0,0,0) &\cropen{-3pt}
&   &     &&     &&    && (-1,0,2,0) &\cropen{-3pt}
&   &     &&     &&    && (0,0,3,-1) &\cropen{-3pt}
&   &     &&     &&    && (0,-1,5,-1) &\cr
\noalign{\hrule}}}}
\newpage
\baselineskip=12pt plus 1pt minus 1pt
\immediate\closeout\referencewrite
\referenceopenfalse
\line{\bf\hfil References\hfil}\vskip.2truein
\input referenc.texauxil
\bye